\def\ts     {\thinspace}
\def\kms    {\ifmmode{{\rm \ts km\ts s}^{-1}}\else{\ts km\ts s$^{-1}$}\fi}
\def\msol   {\ifmmode{{\rm M}_{\odot} }\else{M$_{\odot}$}\fi}
\def\lsol   {\ifmmode{L_{\odot}}\else{$L_{\odot}$}\fi}
\def\lfir   {\ifmmode{L_{\rm FIR}}\else{$L_{\rm FIR}$}\fi}
\def\zsol   {\ifmmode{{\rm Z}_{\odot}}\else{Z$_{\odot}$}\fi}
\def\smghy   {{\rm \ts HS1700.850.1}}

\def\aco    {\ifmmode{{\rm CO}(J\!=\!1\! \to \!0)}\else{{\rm CO}($J$=1$\to$0)}\fi}
\def\bco    {\ifmmode{{\rm CO}(J\!=\!2\! \to \!1)}\else{{\rm CO}($J$=2$\to$1)}\fi}
\def\cco    {\ifmmode{{\rm CO}(J\!=\!3\! \to \!2)}\else{{\rm CO}($J$=3$\to$2)}\fi}
\def\dco    {\ifmmode{{\rm CO}(J\!=\!4\! \to \!3)}\else{{\rm CO}($J$=4$\to$3)}\fi}
\def\eco    {\ifmmode{{\rm CO}(J\!=\!5\! \to \!4)}\else{{\rm CO}($J$=5$\to$4)}\fi}
\def\fco    {\ifmmode{{\rm CO}(J\!=\!6\! \to \!5)}\else{{\rm CO}($J$=6$\to$5)}\fi}
\def\gco    {\ifmmode{{\rm CO}(J\!=\!7\! \to \!6)}\else{{\rm CO}($J$=7$\to$6)}\fi}
\def\hco    {\ifmmode{{\rm CO}(J\!=\!8\! \to \!7)}\else{{\rm CO}($J$=8$\to$7)}\fi}
\def\ico    {\ifmmode{{\rm CO}(J\!=\!9\! \to \!8)}\else{{\rm CO}($J$=9$\to$8)}\fi}
\def\jco    {\ifmmode{{\rm CO}(J\!=\!10\! \to \!9)}\else{{\rm CO}($J$=10$\to$9)}\fi}
\def\kco    {\ifmmode{{\rm CO}(J\!=\!11\! \to \!10)}\else{{\rm CO}($J$=11$\to$10)}\fi}
\def\ci     {\ifmmode{{\rm C}{\rm \small I}}\else{C\ts {\scriptsize I}}\fi}
\def\hi     {\ifmmode{{\rm H}{\rm \small I}}\else{H\ts {\scriptsize I}}\fi}
\def\hh     {\ifmmode{{\rm H}_2}\else{H$_2$}\fi}
\def\cone {\ifmmode{{\rm C}{\rm \small I}(^3\!P_1\!\to^3\!P_0)}
     \else{C\ts {\scriptsize I}{\small$(^3\!P_1\!\to^3\!\!\!P_0)$}}\fi}
\def\ctwo {\ifmmode{{\rm C}{\rm \small I}(^3\!P_2\!\to^3\!P_1)}
     \else{C\ts {\scriptsize I}{\small$(^3\!P_2\!\to^3\!\!\!P_1)$}}\fi}
\def\cij {\ifmmode{{\rm C}{\rm \small I}\,(^3P_i\to^3P_j)}\else{C\ts {\scriptsize I}\,{\small$(^3P_i\to^3P_j)$}}\fi}
\def\cii    {\ifmmode{{\rm C}{\rm \small II}}\else{C\ts {\scriptsize II}}\fi}
\def\tex {\ifmmode{{T}_{\rm ex}}\else{$T_{\rm ex}$}\fi}
\def\tmb {\ifmmode{{T}_{\rm mb}}\else{$T_{\rm mb}$}\fi}
\def\tkin {\ifmmode{{T}_{\rm kin}}\else{$T_{\rm kin}$}\fi}
\def\microns {\ifmmode{\mu{\rm m}}\else{$\mu$m}\fi}
\def\um{\ifmmode{\mu{\rm m}}\else{$\mu$m}\fi}
\def\nhh   {\ifmmode{n({\rm H}_2)}\else{$n$(H$_2$)}\fi}
\def\gradv {\ifmmode{{\rm dv/dr}}\else{dv/dr}\fi}

\def\smghy   {{\rm  HS1700.850.1}}

\documentclass[usenatbib,useAMS]{mn2e}
\usepackage{graphicx}
\usepackage{epstopdf}
\usepackage{ctable}
\usepackage{url}
\usepackage{times}

\newcommand{\apj}{ApJ}

\newcommand{\mnras}{MNRAS}

\newcommand{\acknowledgments}{\begin{small}\section*{Acknowledgments}\end{small}}

\title[A blind $^{12}$CO redshift search]{A millimeter-wave redshift search for  the unlensed HyLIRG, HS1700.850.1} 

\author[S.~C.\ Chapman et al.]{
\parbox[t]{\textwidth}{
S.\,C.\ Chapman$^{1}$\thanks{scott.chapman@dal.ca},
F.\ Bertoldi,$^{2}$
Ian Smail,$^{3}$
C.\,C.\ Steidel,$^{4}$
A.\,W.\ Blain,$^{5}$
J.\,E.\ Geach,$^{6}$
M.~Gurwell,$^{7}$ 
R.\,J.\ Ivison,$^{8,9}$
G.\,R.\ Petitpas,$^{7}$ 
N.\ Reddy$^{10}$}
\vspace*{6pt} \\
$^1$Department of Physics and Atmospheric Science, Dalhousie University, Halifax, NS, B3H 4R2, Canada\\
$^2$ Argelander-Institute of Astronomy, Bonn University, Auf dem Hugel 71, D-53121 Bonn, Germany\\
$^{3}$Centre for Extragalactic Astronomy, Department of Physics, Durham University, South Road, Durham DH1 3LE\\
$^{4}$Cahill Center for Astronomy and Astrophysics, California Institute of Technology, MS 249-17, Pasadena, CA 91125, USA\\
$^{5}$Physics and Astronomy, University of Leicester, University Road Leicester, LE1 7RH\\
$^{6}$Centre for Astrophysics Research	, Science \& Technology Research
Institute, University of Hertfordshire, Hatfield AL10 9AB, UK\\
$^{7}$Harvard-Smithsonian Center for Astrophysics, 60 Garden Street, Cambridge, MA 02138\\
$^8$Institute for Astronomy, University of Edinburgh, Royal Observatory, Blackford Hill, Edinburgh, EH9 3HJ, UK\\
$^9$European Southern Observatory, Karl Schwarzschild Strasse 2, D-85748 Garching, Germany\\
$^{10}$Department of Physics and Astronomy, UC Riverside, 900 University
Avenue, Riverside, CA 92521
}
\usepackage{hyperref}

\begin{document}

%\date{Accepted ???. Received ???; in original form ???}
\date{Submitted for publication in MNRAS}

\pagerange{\pageref{firstpage}--\pageref{lastpage}} \pubyear{2013}

\maketitle

\label{firstpage}

\begin{abstract}
We report the redshift of an unlensed, highly obscured submillimetre galaxy
(SMG), \smghy, the brightest  SMG (S$_{\rm 850\mu m}$=19.1\,mJy) detected in the JCMT/SCUBA-2 Baryonic Structure Survey,
based on the detection of its $^{12}$CO line emission.  
Using the
IRAM PdBI-WIDEX with 3.6\,GHz band width, we serendipitously detect an emission line at 150.6\,GHz.
From a search over 14.5\,GHz in the 3-mm and 2-mm atmospheric windows, we confirm the identification of this line as $^{12}$CO(5--4) at $z=2.816$, meaning that it does not reside in the $z\sim2.30$ proto-cluster in this field.  
Measurement of the 870$\mu$m source size ($<$0.85$''$) from the Sub-Millimeter Array (SMA)  confirms a  compact emission in a S$_{\rm 870\mu m}=$14.5\,mJy, L$_{IR}$$\sim$10$^{13}$\,L$_\odot$ component, suggesting an Eddington-limited starburst.
We use the double-peaked $^{12}$CO line profile measurements along with the SMA size constraints to study 
the gas dynamics of a HyLIRG, estimating the gas and dynamical masses of \smghy. 
While \smghy\ is one of the most extreme galaxies known in the Universe,
we find that it occupies a relative void in the Lyman-Break Galaxy distribution in this field.  
Comparison with other extreme objects at similar epochs (HyLIRG Quasars), and cosmological simulations, suggests such an {\it anti-bias} of bright SMGs could be relatively common, with the brightest SMGs rarely occupying the most overdense regions at $z=$2--4.
\end{abstract}


\section{Introduction} \label{S:intro}

\begin{figure*}
\centering
\includegraphics[width=19.0cm,angle=0]{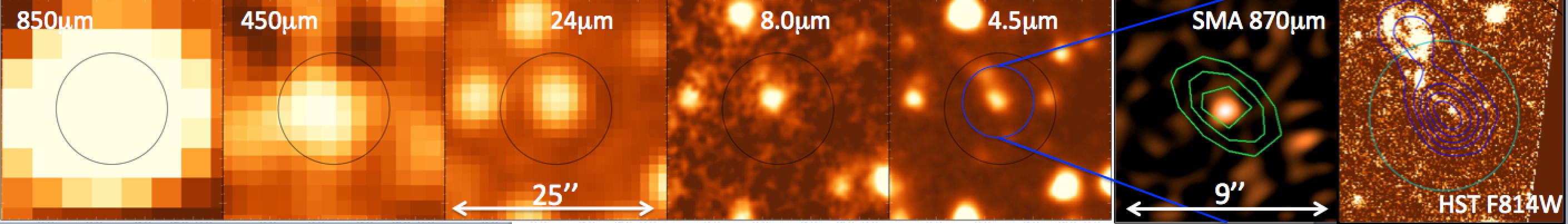}
\caption{
{\bf Left to right:} 
Five cutout images 25$''$ on a side  of the longer wavelength data for HS1700.850.1 
(850$\mu$m, 450$\mu$m, 24$\mu$m, 8$\mu$m, 4.5$\mu$m), highlighting the likely identification of the SMG with an IR-luminous component (black circle shows the 15$''$ SCUBA-2 beam size).
{\bf Second from right:} 
A 9$''$$\times$9$''$ zoom in from the SMA compact+extended configuration 870$\mu$m map, with a $\sim0.74''$ synthesized beam with PdBI $^{12}$CO(5--4) contours overlaid. Most of  the SCUBA-2 flux (14.5 mJy) is recovered in a single compact component, unresolved with a 1$''$ beamsize. No other significant sources are detected within the 15$''$ diameter of the SCUBA-2 beam. Further, the compact configuration data on its own measures a flux of 19$\pm$3 mJy, comparable to the SCUBA-2 flux, and suggestive that the source may be marginally resolved on scales $>1''$.
{\bf Far right:} 9$''$ cutout of the {\it HST}-ACS F814W image centred on the SMA detection, with IRAC ch-2 contours overlaid, shows  HS1700.850.1  to be well identified with a faint tad-pole like $I_{AB}\sim26$ galaxy, which is marginally detected in ground based imaging (${\cal R}=26.1$).  
}
\label{spectra} 
\end{figure*}

Blank-field mm and submm continuum surveys have
discovered hundreds of dusty, star-forming submm galaxies (SMGs) over the past decade
({e.g., Smail et al.\ 2002; Borys et al.\ 2003; Greve et al.\ 2004; Coppin et al.\ 2006; Bertoldi et al.\ 2007; Scott et al.\ 2008; Weiss et al.\ 2009}), while
{\it Herschel}-SPIRE has mapped close to 10$^3$ deg$^2$ in the 500, 350, 250$\mu$m bands, leading to $\sim$10$^4$ SMGs being detected (e.g., Oliver et al. 2010). At higher flux limits SPT has mapped 2500 deg$^2$ (e.g., Mocanu et al.\ 2013), and {\it Planck} has mapped the entire sky (Planck collaboration et al.\ 2014), 
finding many extreme and/or  gravitationally lensed  SMGs.
The search for the most luminous star forming galaxies from 100's of deg$^2$ in the Universe has ensued (e.g,. Fu et al.\ 2013, Ivison et al.\ 2013). However some of the SMGs discovered in early $\sim$100 arcmin$^2$ surveys remain the intrinsically brightest, most extreme known (e.g., SMM\,J02399 -- Ivison et al.\ 1998, 2010, GN20 -- Pope et al.\ 2006; 
Younger et al.\ 2008, and COSMOS-AzTEC1 -- Scott et al.\ 2008;
Younger et al.\ 2010).

Determining the redshifts of unlensed SMGs has largely been a process of using  
weak counterparts in the rest-frame ultraviolet and optical to make spectroscopic determinations (e.g,. Chapman et al.\ 2003, 2005),
or combining near- and mid-IR photometry to obtain reliable photometric redshift estimates (e.g., Wardlow et al.\ 2011; Simpson et al.\ 2014).
However, because the high extinction of SMGs means that they often have only very faint (if any) counterparts in the optical/near-IR bands, the spectroscopic and even photometric redshift distributions have remained incomplete and biased (e.g., Simpson et al.\ 2014).

The largest spectroscopic SMG redshift survey to date was based on radio-identified SMGs
(Chapman et al.\ 2005).
The radio identification is sometimes inaccurate (Hodge et al.\ 2013) and may bias the redshift distribution
since radio emission may remain undetected even in the deepest radio
maps for sources at $z>3$. 
Optical spectroscopic followup to ALMA-identified SMGs (Simpson et al.\ 2014; A.\ Danielson et al.\ in prep) have removed the identification bias, yet cannot remove the spectroscopic bias. 

An alternative route to determine the redshift of an SMG is
through observations of $^{12}$CO emission lines at cm or mm
wavelengths. The $^{12}$CO lines arise from the molecular gas, the fuel for star
formation, and can thus be related unambiguously to the submm
continuum source. Therefore, these observations do not require any
additional multi-wavelength identification and circumvent many
of the problems inherent to optical spectroscopy of SMGs. 
However, to date the sensitivity of mm-wave facilities has limited the approach to followup 
of bright, gravitationally lensed SMGs (e.g., Weiss et al.\ 2009; Swinbank et al.\ 2010; Harris et al.\ 2012; Vieira et al.\ 2013;  Zavala et al.\ 2015).

While the  IRAM Plateau de Bure Interferometer (PdBI)  is  sensitive enough to detect $^{12}$CO in unlensed SMGs (e.g. Bothwell et al.\ 2013), its 3.6\,GHz bandwidth receivers limit  
 this approach due to the large time requirement for blind searches of $^{12}$CO lines in redshift space via multiple frequency tunings.  
Walter et al.\ (2012) observed the HDF-N field with the PdBI over the full 3mm window to a similar depth as the Bothwell et al.\ (2013) survey, requiring 110\,hrs on-source. In addition to blindly detecting two faint $^{12}$CO emitting galaxies, they  identified the redshift of the S$_{\rm 850\mu m}\sim$5\,mJy HDF850.1 as $z=5.3$. 
The 8-GHz instantaneous,
dual-polarization bandwidth of ALMA, coupled with its sensitivity, improves this approach, but still makes it a relatively expensive prospect to blindly obtain $^{12}$CO redshifts for even the brightest SMGs with S$_{\rm 850\mu m}\sim$8\,mJy.

In this paper, we describe a photometric redshift guided,  mm-wave search for $^{12}$CO line emission from \smghy, one of the brightest unlensed SMGs ever discovered,
with  $S_{850\micron}=19.1$\,mJy. The  target field was chosen from the 24 survey fields of the Keck Baryonic Structure Survey (KBSS -- e.g., Rudie et al.\ 2012).  Several of these fields show strong galaxy overdensities (e.g., Steidel et al.\ 2000, 2005, 2010), and together represent a complete, unbiased census of  overdensities in a well-calibrated field galaxy survey at $z>2$.  %Six of these fields were followed up with SCUBA-2 on the JCMT, including the two fields hosting massive protoclusters. 
The HS1700+64 field in which \smghy\ was found hosts a well characterized $z=2.30$ protocluster, studied in detail in several works (Steidel et al.\ 2005; Shapley et al.\ 2005; %ApJ...626..698S
Digby-North et al.\ 2010). %MNRAS.407..846D, ). 
The  faintness of \smghy\  at near-IR/optical wavelengths precluded spectroscopic observations, despite lying in this well studied field.

We use cosmological parameters $\Omega_m$ = 0.3, $\Lambda$ = 0.7, and H = 70 km s$^{-1}$ Mpc$^{-1}$ throughout the paper; at $z=2.82$, this corresponds to an angular scale of 7.94 kpc arcsec$^{-1}$, or 0.48 Mpc arcmin$^{-1}$.

\begin{figure*}
\centering
\includegraphics[width=15.0cm,angle=0]{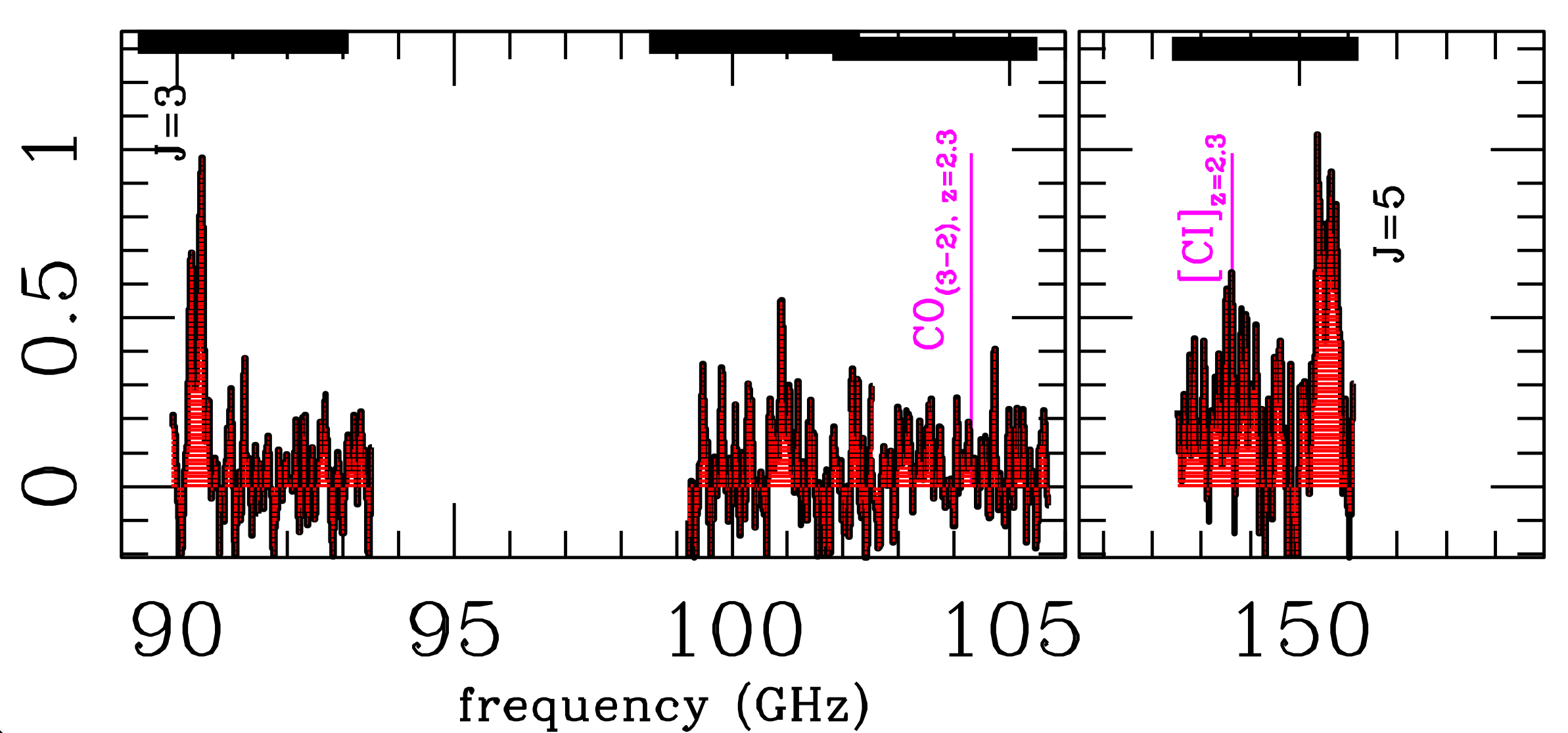} 
\caption{A composite of all spectral scans towards \smghy, shown at a velocity resolution of 60\,\kms\ (arbitrary flux scaling). $^{12}$CO emission
lines are seen at $\sim90.2$ and $150.6$\,GHz, the latter corresponding to $^{12}$CO(5--4). The originally targeted $z=2.31$ $^{12}$CO  and [CI] line frequencies  are highlighted (magenta), with a curious excess seen at the [CI] frequency. The four tuning bands are marked with heavy lines, showing a slight overlap near 101\,GHz.
 }
\label{3mm-scan} 
\end{figure*}

\section{Observations and strategy \label{observations}}

The HS1700+64  field was targeted with SCUBA-2 at the JCMT as part of a submm-wave followup of six KBSS fields (K.\ Lacaille et al.\ in prep), covering $\sim0.19$ deg$^2$.
The brightest source in this 850$\mu$m survey  is  \smghy\ with S$_{850\mu m}$=19.1$\pm$0.8\,mJy.
Fig.\ 1 shows  cutouts of \smghy\ at SCUBA-2 and {\it Spitzer} wavelengths, revealing  a likely identification at 4.5$\mu$m. 

The HS1700+64 field contains a massive proto-cluster at $z=2.30$ (Steidel et al.\ 2005), and given its 1.5$'$ proximity to the cluster centre, \smghy\ was considered a likely cluster member.  \smghy\ has  two candidate {\it Spitzer}-IRAC identifications within 4$''$ of the 850$\mu$m centroid, one of which is associated with a well detected complex of $z=2.305$ galaxies. %
IRAM-PdBI observations targeting $^{12}$CO(3--2) and [CI] at $z=2.305$ in the 3mm and 2mm bands respectively  showed no detectable line emission, indicating this redshift was incorrect (Fig.~2).
However, the WIDEX receiver's 3.6\,GHz wide spectrum %in each band 
revealed a strong line detected serendipitously 
at  150.6\,GHz near the edge of the chosen 2mm-band setting (Figs.~2,3). Two followup  observations targeted the three most likely redshifts, assuming this 150.6\,GHz line was a $^{12}$CO rotational transition, as described in \S~3.1. 
We also draw on  {\it Spitzer}-IRAC, ground-based and {\it HST} optical datasets described  by Steidel et al.\ (2011).  All flux measurements are shown in Fig.~4, and discussed in \S~3.1.

\begin{figure}
\centering
\includegraphics[width=8.5cm,angle=0]{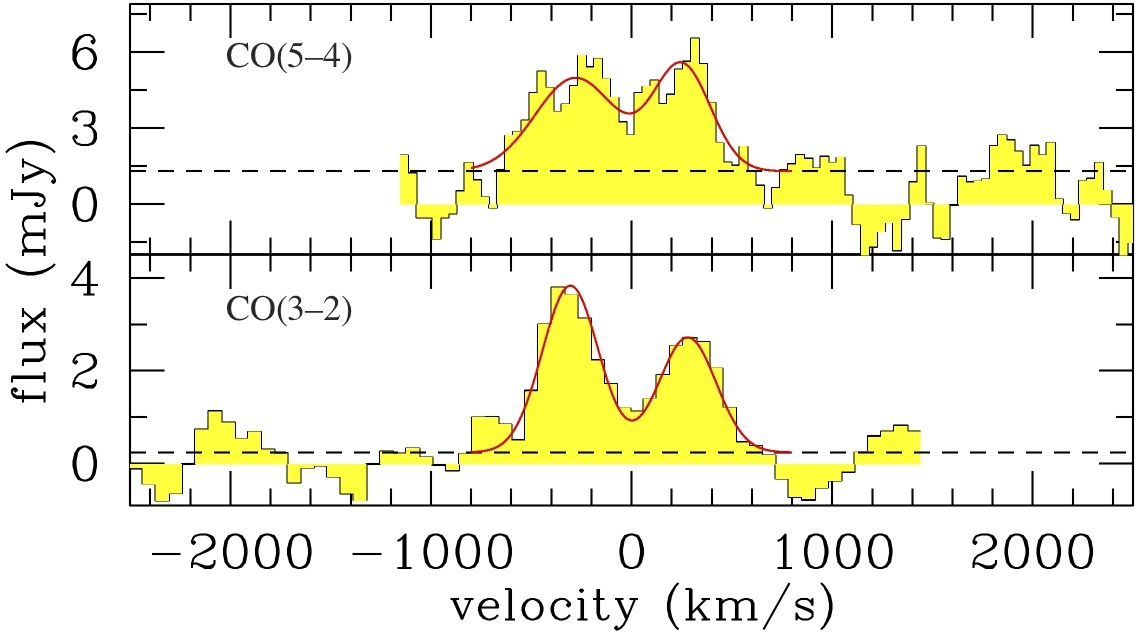} 
\caption{Spectra of the $^{12}$CO(3--2) and $^{12}$CO(5--4)  lines towards \smghy. The spectral resolution is 60\,\kms\ for
both lines. %See Table\,\ref{colines} for the fit parameters. 
We find only marginal evidence for any spatial separation in the two peaks of the CO lines $<1''$.
Double Gaussian fits to each line profile are overlaid, emphasizing the $\sim$50\% higher  peak flux  in the blue component of the CO(3--2) profile.
 Line ratio and dynamical modelling issues are discussed in detail in the text.
 }
\label{3mm-scan} 
\end{figure}

\begin{table}
\label{tab:par}
 \flushleft{
  \caption{PdBI observations log.}
\begin{tabular}{llccccccc}
\hline
Run-ID  &  center freq &  t$_{\rm int}^a$ & Depth$^b$  &  obs date & beamsize \\ %   L$'$(CO) & $\sigma$(CO) & z(UV) &  z(CO)   & comments \\   % & velocity (channel peak)
{}  & {(GHz)} &  {(hr)}  & (mJy) & {} \\ %& (L$_\odot$) & (km/s)  &  &    &  \\   % & velocity (channel peak)
%     3mmflux broadline?
\hline
S14ck001  & 103.8	   & 3.34 & 0.60 & 09/06/14 & 5.1$''$$\times$4.2$''$ \\ %
S14ck002  & 149.2	   & 3.45 & 1.12 & 11/06/14 & 3.5$''$$\times$2.7$''$ \\%
D14d001   & 100.8	   & 1.50 & 0.95 & 28/07/14 & 4.7$''$$\times$4.0$''$ \\% in map, noise seems higher
D14d002   &  91.7	   & 3.15 & 0.43 & 04/08/14 & 4.8$''$$\times$4.1$''$ \\%
\hline
\hline
%C4nb    &    0.290  2.8   2.846 \\  %3005
%C6bx    &    0.114  2.8   2.844 \\  %2810
%\hline
%C1 & 15:51:50.80      +19:11:40.0 & 0.313 & 5.1 &&& n/a & 2.853 & no known optical counterpart \\  %further than primary beam.
\end{tabular}
} 
\medskip 
\flushleft{$^a$  5-Antenna  integration. }
\flushleft{$^b$  per 50MHz channel. }
%\flushleft{$^a$  6-Antenna equivalent integration. }
\end{table}

%% in d002 for z=2.82, 0.041mJy RMS in 180ch => 0.55mJy RMS in 1ch = 55km/s
%% in d001 for z=3.6, 0.130mJy RMS in 180ch => 1.7.mJy RMS in 1ch = 55km/s
%% in s14ck001 for z=2.3, 0.077mJy RMS in 180ch => 1.0mJy RMS in 1ch = 55km/s
%% in s14ck002 for z=2.3 CI, 0.150mJy /RMS in 180ch => 1.34mJy RMS in 1ch = 55km/s

%\begin{table}
%\label{tab:par}
% \flushleft{
%  \caption{source positions}  %all the same so ignore
%\begin{tabular}{llccccccc}
%\hline
%source &	RA	Dec \\
%CO(3-2)  & 17:01:17.649	+64:14:37.28 \\
%continuum 90GHz & 17:01:17.779 +64:14:37.85
%cont 104GHz & 17:01:17.607 +64:14:37.80 \\
%2mm ctm & 17:01:17.649 +64:14:37.86 \\ 
%co54 & 17:01:17.676 +64:14:37.64 \\
%\hline
%\end{tabular}
%} 
%\end{table}

\begin{table}
\label{tab:par}
 \flushleft{
  \caption{Photometry of HS1700.850.1}
\begin{tabular}{llccccccc}
\hline
Band  &  Flux  & Instrument\\ %  Flux GN20 & Flux AzTEC8   \\ %   L$'$(CO) & $\sigma$(CO) & z(UV) &  z(CO)   & comments \\   % & velocity (channel peak)
%{}  & {}  \\ %&  {(hr)}  & (mJy) & {} \\ %& (L$_\odot$) & (km/s)  &  &    &  \\   % & velocity (channel peak)
%     3mmflux broadline?
\hline
91.7GHz  & 0.25$\pm$0.07 (mJy) & PdBI \\%	   & 0.33$\pm$0.07 &{}  \\ % line free flux (ignoring CO32 ... with it's 0.37mJy
100.8GHz  & 0.31$\pm$0.13 (mJy) & PdBI\\%	   & {} & {} \\ %
103.8GHz   & 0.38$\pm$0.08 (mJy) & PdBI\\%	   & {} & {}\\ %
149.2GHz   &  1.39$\pm$0.15 (mJy)	& PdBI\\%   & 0.9$\pm$0.2  & {} \\ %linefree, without  co54.... with it's 1.55mJy
%220GHz  {} & mambo &  &\\
%272GHz {} & aztec & & 5.5$\pm$1.3\\
850$\mu$m & 19.1$\pm$0.8 (mJy) & SCUBA-2  \\ %& 17.1$\pm$1.3 & 15.7$\pm$1.6 \\  
870$\mu$m  & 14.5$\pm$1.1  (mJy) & SMA \\ % & 21.3$\pm$1.1 & 19.7$\pm$1.3 \\  
450$\mu$m  & 45$\pm$6 (mJy) & SCUBA-2\\ % & 45 & 45 \\
24$\mu$m  & 171$\pm$7 ($\mu$Jy)  & MIPS\\ % & 45 & 45 \\
8.0$\mu$m  & 13.2$\pm$1.2 ($\mu$Jy) & IRAC \\ % & 45 & 45 \\
4.5$\mu$m IRAC & 10.2$\pm$0.9 ($\mu$Jy) & IRAC \\ % & 45 & 45 \\
$K_s$(2.18$\mu$m)  & 23.20$\pm$0.13 (AB mag) & P200/WIRC\\ % & 45 & 45 \\
$J$(1.25$\mu$m)  & 24.4$\pm$0.2 (AB mag) & P200/WIRC \\ % & 45 & 45 \\
$I$(F814W)  & 25.8$\pm$0.2 (AB mag) & HST/ACS \\ % & 45 & 45 \\
$ {\cal R}$ $^a$ & 26.1$\pm$0.3 (AB mag) & WHT/PFC \\ % & 45 & 45 \\
$g'$  & 27.0$\pm$0.3 (AB mag) & WHT/PFC\\ % & 45 & 45 \\
$u'$  & $>$27.1, 3$\sigma$ (AB mag) & WHT/PFC  \\ % & 45 & 45 \\

\hline
\hline
\end{tabular}
} 
\medskip 
\flushleft{$^a$ The $ {\cal R}$ filter is centered at 6900\AA\ and thus redder than the standard $R$ -- close to  mid-way between SDSS $r'$ and $i'$  in  effective wavelength }
\end{table}

\subsection{JCMT/SCUBA-2 Observations}

SCUBA-2 observations  were conducted in Band 2 weather conditions ($\tau_{\rm 225 GHz} < 0.08$) over 6 nights between 23rd May and 15th Oct.\ 2013, totalling 15 hours of on-sky integration in individual 30 min scans. A standard 3$'$ diameter ``daisy'' mapping pattern was used, which keeps the pointing centre on one of the four SCUBA-2 sub-arrays at all times during exposure. 
Data reduction followed standard recipes (e.g., Geach et al.\ 2013), with individual 30 min scans  reduced using the dynamic iterative map-maker of the SMURF package (Jenness et al.\ 2011), flat-fielding, and then solving for model components assumed to make up the bolometer signals (atmospheric, astronomical, and noise terms).
The signal from each bolometer's time stream is then re-gridded onto a map, with the contribution to a given pixel weighted according to its time-domain variance.
Filtering of the time-series is performed in the frequency domain, with band-pass filters equivalent to angular scales of $2'' < \theta < 120''$, along with 10$\sigma$ deviation {\it spike} removal in a moving boxcar, and DC step corrections. Finally, since we are interested in (generally faint) extragalactic point sources, we apply a beam matched filter to improve point source detectability, resulting in a map that is convolved with an estimate of the 850$\mu$m beam. 
The map reaches  central noise levels of $\sigma_{850}$=0.51\,mJy and $\sigma_{450}$=6.4\,mJy.

\subsection{IRAM/PdBI Observations}

The IRAM-PdBI  observations were taken in D configuration using the  5-antenna sub-array  (properties of the observing runs and resulting synthesized beam sizes are summarized in Table~1), at a
%centered at  frequencies of 91.7, 100.8, 103.8, and 149.2\,GHz  at a 
pointing center: RA=17 01 17.779, Dec=+64 14 37.85, J2000.0, taken from the SCUBA-2 centroid.
Flux calibration was achieved by observing various calibrators (3C273, 3C345, B0234+285, B1749+096). The quasar B1300+580 was used as a phase and amplitude calibrator. Data were processed using the most recent version of the {\sc GILDAS} software. 
We resampled the cubes in 90 km s$^{-1}$ channels, and imaged them using the GILDAS suite {\it mapping}, adopting natural weighting. The full PdBI spectrum is shown in Fig.~2.
To obtain flux measurements, we deconvolved the visibilities using the CLEAN task with natural weighting, and applied the corresponding primary beam correction (negligible at the small 3$''$ offset of the \smghy\ position).

\subsection{SMA observations}

Followup SMA observations were performed on  August 15  and September 7, 2014 in the compact and extended 
configurations respectively (beam sizes $\sim$2$''$ and $\sim$0.85$''$ respectively) in good weather ($\tau_{\rm 225GHz}\sim 0.08$) 
with a total on-source integration time of approximately 6\,hr. The USB was tuned to 345 GHz, and 
combined with the LSB for an effective bandwidth of $\sim 4$ GHz at 340 GHz, which yielded a final %synthesised image 
rms of 1.1\,mJy.  The pointing centre was the same as  that for the PdBI observations described above.
The data were
calibrated using the {\sc mir} software package (Scoville 1993), modified for the SMA.  Passband calibration was done using 3C\,84,
3C\,111, and Callisto.  The absolute flux scale was set using observations of Callisto and is estimated to be accurate to better than 20\%.  Time-dependent complex gain calibration was
done using 1858+655 (0.6\,Jy, 21.8$^{\circ}$ away) and 1827+390
(1.8\,Jy, 37.7$^{\circ}$ away).  
We detect \smghy\ in the synthesized image at $\sim 13\sigma$.  The
calibrated visibilities were best fit by a single point source 
with an integrated flux density of $S_{\rm 890\mu m} = 14.5\pm1.1$\,mJy at a position of
$\rm\alpha(\textrm{J2000}) =$17 01 17.779 and
$\rm\delta(\textrm{J2000}) =$ +64 14 37.85, 
consistent within 0.3$''$ with the centroid of the line and continuum measurements from PdBI.  
The astrometric
uncertainties are $\Delta\alpha=0.24$\,arcsec (0.20\,arcsec
systematic; 0.13\,arcsec statistical) and $\Delta\delta=0.22$\,arcsec
(0.19\,arcsec systematic; 0.10\,arcsec statistical).

\section{Results}

\subsection{The redshift search}
Our original IRAM-PdBI observations  searched for the  $^{12}$CO and [CI] lines in  \smghy\ at a redshift $\sim2.30$.  This was motivated by  it lying near the core of a massive $z=2.31$ protocluster in the  Keck Baryonic Structure Survey (Steidel et al. 2011), and  having a plausible 4.5$\mu$m identification with a complex of known $z=2.305$ BX galaxies. Being such an extreme source we would not expect to find it by chance in the small 0.03 deg$^2$ H1700+64 pointing, or even in the 0.19 deg$^2$ survey of the six KBSS fields without the high density environment of the proto-cluster in this field. Using typical single dish 850$\mu$m counts (Coppin et al.\ 2006; Weiss et al.\ 2009) we expect a surface density of just 0.01 deg$^{-2}$ for such 19\,mJy sources. %With knowledge that 
The source counts of the brightest submm sources (S870$\mu$m$>$9\,mJy) measured by ALMA fall more steeply than those measured with lower resolution single-dish observations due to the multiplicity of the brightest sources when observed at arcsecond resolutions (e.g., Karim et al.\ 2013). 
As \smghy\  retains at least $\sim$15\,mJy in the interferometric followup (and as much as 19\,mJy if the compact configuration SMA data is taken at face value), the expected $>$15\,mJy source density  drops by another factor of ten based on the ALMA-derived count of Karim et al.\ (2013) or Simpson et al.\ (2015b).  

The IRAM-PdBI observations  showed no such lines at the expected frequencies, and thereby implied that the proposed identification and redshift of $z\sim2.31$ was incorrect. 
As the 7.2\,GHz of frequency coverage did reveal a strong line detected serendipitously at  150.6 GHz near the edge of the 2mm setting, we began a search to identify the correct redshift.  The centroid of this 150.6\,GHz line detection, as well as the PdBI 2mm and 3mm continuum detections of \smghy, both localize the emission from the source within optical/IR imagery to within $\sim0.5''$ (Fig.\ 1), while followup SMA 870$\mu$m continuum imaging unequivocally associates the 850$\mu$m SCUBA-2 source to a single IRAC galaxy which is well detected in the {\it HST} F814W image (Fig.~1) revealing a faint disturbed, tadpole-like galaxy. 
Ground-based imagery marginally detects the galaxy at  $ {\cal R} =26.1$, $g=27$, $U>27.1$, making optical spectroscopy very difficult, and the $z=2.82$ redshift would be difficult for near-IR spectroscopy, although the $K_{s, AB}=23.2$ is feasible in general with a 10m telescope. Note that these colors (e.g., $g'-{\cal R}$ = 0.9 mag) make sense for a heavily reddened object at $z$=2.82. 
However, the detection at $g'$-band, places loose constraints on the redshift to $z<4$, while the comparable IRAC fluxes,  S(4.5$\mu$m)=10.2$\mu$Jy  .and S(8.0$\mu$m)=13.2$\mu$Jy  suggests the 1.6$\mu$m stellar bump should lie between, ranging between 2.4$<$$z$$<$3.7.  
The very red $K_s-4.5\mu$m coloyr (1.8 mags AB) 
 is  large, though consistent with a high stellar mass object. 
The MIPS 24$\mu$m flux of 0.17\,mJy compared to the S$_{\rm 850 \mu m}$ flux suggests that  PAH emission is not contributing to the 24$\mu$m band, pushing the redshift to $z>2.5$, as illustrated in the SED template fit (Fig.~4).
Our final constraint on the redshift comes from the long-wavelength photometry (Fig.~4, Table~2), where 3mm, 2mm, 870$\mu$m, and 450$\mu$m fluxes suggest a similar range of 2.5$<$$z$$<$4.5 for a likely range of SMG dust temperatures, 25\,K$<$${\rm T_d}$$<$45\,K.

While these constraints may not seem very useful, the fact that we have a solid line detection at 150.6\,GHz means that only three possible $^{12}$CO identifications provide redshifts in this range: $^{12}$CO(4--3) at $z=2.82$,  $^{12}$CO(5--4) at $z=3.58$, or $^{12}$CO(6--5) at $z=4.37$. Aided by these photometric redshift estimates, in two subsequent  PdBI followup observations we searched for $z=3.58$  $^{12}$CO(4--3) at 100.4 GHz,  and then at 91.5 GHz searched simultaneously for $z=2.82$ $^{12}$CO(4--3) and [CI] at $z=4.37$. With a detection of a second line (Fig.\ 2) we confirmed the redshift $z$=2.816 for one of the most intrinsically luminous galaxies in the distant Universe ($L_{\rm FIR}\sim 5\times 10^{13} L_\odot$, \S~3.3)
and secured a second $^{12}$CO line for astrophysical diagnostics.
Blind $^{12}$CO redshifts for unlensed SMGs are still exceedingly rare in the literature. We are only aware of Daddi  et al.\ (2009) GN20 at  $z=4.05$ (a truly serendipitous discovery),  and  the $z=5.2$ for HDF850.1 (Walter et al.\ 2012).

%%%%%

\begin{figure}
\centering
\includegraphics[width=8.5cm,angle=0]{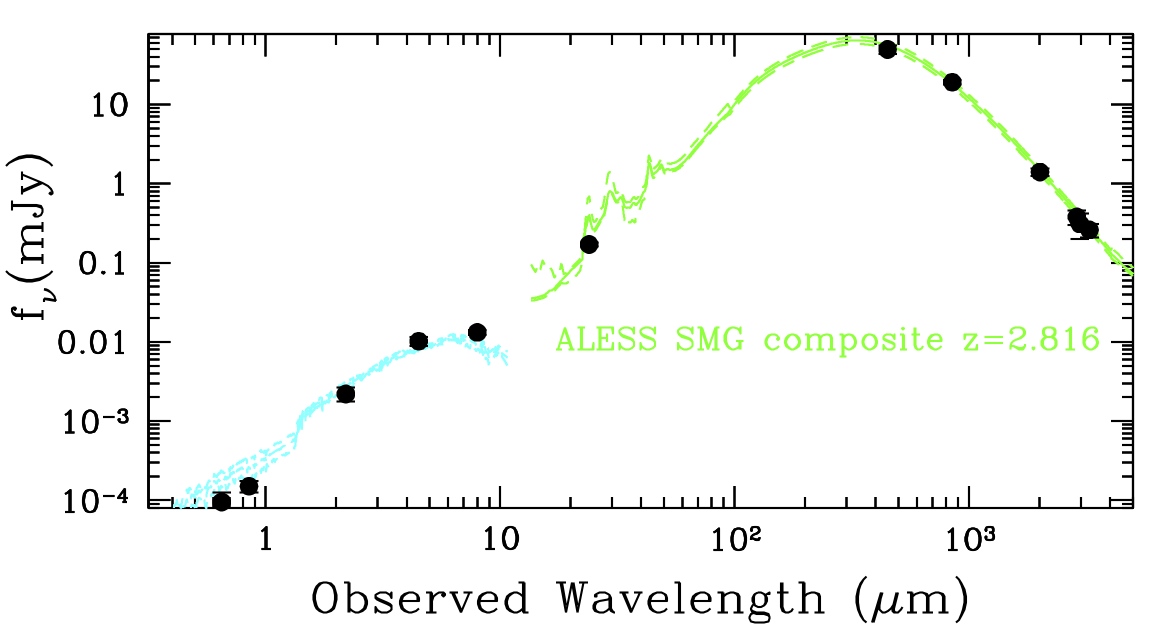} 
\caption{The SED of \smghy, with the Swinbank et al.~(2014) composite SMG template overlaid. Significantly hotter or cooler SEDs than this $\sim35K$ template are reasonably rejected by the photometry. The template is split at a rest frame 10$\mu$m in order to fit the optical/near-IR faint component as decoupled from the luminous far-IR to submm dust component. 
The far-IR luminosity of this model is $3.8\cdot10^{13}\,\lsol$. 
}
\label{dustsed} 
\end{figure}

\begin{table}
      \caption[]{Gaussian decomposition of the  CO line  spectra. Values for the total line are estimated from integrating the continuum-subtracted spectra, while the red and blue component properties are estimated from the Gaussian fits. Errors in redshift estimation are 
         d$z$=0.001. The measured FWHM for the total line is quoted as  $\sigma$, 2.355$\times$ smaller.
$L'CO$ is in units of 10$^{10}$ K km s$^{-1}$ pc$^{2}$.
         Note that for $\alpha$(CO)=1 M$_\odot$ (K km s$^{-1}$ pc$^{2}$)$^{-1}$,
                  as estimated in Bothwell et al.\ (2013), $L'CO_{\rm 1-0}$ = M(gas), which should then be corrected by an additional 15\% for Helium.
         }
         \label{decomp}
          \begin{tabular}{l c c c}
            \hline
            \noalign{\smallskip}
            & blue  & red  &  total \\
            \noalign{\smallskip}
            \hline
            \noalign{\smallskip}           
         $z_{{\rm (3-2)}}$ & $2.812$ & $2.820$ & 2.816\\
         $\delta$v(3-2) [\kms]$^{a}$ &$-305\pm9$ &$+280\pm11$ & 0\\ 
         $\delta$v(5-4) [\kms]$^{a}$ &$-280\pm21$ &$+250\pm19$ & 0\\ 
         $\sigma$(3-2) [\kms] &$140\pm25$ & $142\pm20$ & $400\pm57$$^c$\\
         $\sigma$(5-4) [\kms] &$200\pm45$ & $143\pm24$ & $349\pm45$$^c$\\
%         $S_{\nu\,{\rm 3-2}}$ & $0.58\pm0.18$ & 1 & 999 \\
%         $S_{\nu\,{\rm 5-4}}$ & $0.58\pm0.18$ & 1 & 999 \\
         $I_{\rm 3-2}$ [Jy km s$^{-1}$] & $1.26\pm0.16$ & $0.88\pm0.13$ &  $2.32\pm0.21$  \\ %(2.6 inc ctm)
         $I_{\rm 5-4}$ [Jy km s$^{-1}$] & $1.84\pm0.23$ & $1.5\pm0.21$ & $3.44\pm0.30$ \\ %(5.1 inc ctm)
	$L'CO_{{\rm 3-2}}$ & $5.18$ & 3.62 & 9.65 \\
         $L'CO_{{\rm 5-4}}$ & $2.72$ & 2.21 & 5.08 \\
         $L'CO_{{\rm 1-0}}\,^b$ & $9.95$ & 6.95 & 18.56 \\
%         $log(M$_{\rm gas})\,^c$ & $9.95$ & 6.95 & 18.56 \\
                     \noalign{\smallskip}
            \hline
           \end{tabular}
\begin{list}{}{}
\item[$^{\mathrm{a}}$] Center velocity relative to $z$=2.816.
\item[$^{\mathrm{b}}$] Derived using average r$_{31} = 0.32$ scaling relation of Bothwell et al.\ (2013).
\item[$^{\mathrm{c}}$] These are the best fit single Gaussian to the entire line profile.
\end{list}
\end{table}

\subsection{Source properties}

\smghy\ is very bright at 850$\mu$m, but shows no sign of lensing -- the nearest galaxy detected is the low-mass star forming galaxy BX951 2.9$''$ away ($z$=2.3053, $M^*=2\times10^9$ M$_\odot$). \smghy\ is thus not suffering from differential lensing and associated modelling issues and biases that could affect our interpretation of the $^{12}$CO lines.
The $^{12}$CO  spectra are shown zoomed in  at a velocity resolution of
60\,\kms\ in Fig.~3. 
Both lines are detected at high significance (11.3\,$\sigma$ and 6.9\,$\sigma$ for continuum subtracted (3--2) and (5--4) lines respectively, in integrated intensity). The line profiles for
both lines are similar and well described by a double Gaussian
with a peak separations of 585 and 530\,\kms  for (3--2) and (5--4) $^{12}$CO  lines respectively, and individual peak FWHMs of $\sim$330-470\,\kms. While the (3--2) line shows an apparent peak flux asymmetry   compared with the (5--4) peaks, in fact the  flux ratios of the blue and red lines are quite similar, r$_{53, blue}$=0.69$\pm$0.12 and r$_{53, red}$=0.59$\pm$0.09.
The parameters derived from Gaussian fits to
both line profiles are given in Table~3. %\ref{colines}. 
We find only marginal evidence for any spatial separation in the two peaks of either of the $^{12}$CO lines $<1''$.
The frequencies unambiguously identify the lines as CO(3--2) and CO(5--4).
Combining the centroids of both lines, we derive a variance-weighted mean redshift for
\smghy\ of $z=2.816 \pm 0.001$.
While the observed frequencies might also be interpreted as $^{12}$CO(6--5)
and $^{12}$CO(10--9) at $z=6.64$,  
the $^{12}$CO ladder is not equidistant in frequency
which results in small differences for the frequency
separation of the line-pairs as a function of rotational quantum
number. The frequency separation is 58.58 and 58.53\,GHz for
the $^{12}$CO line-pairs at redshifts 2.82 and 6.64 
respectively. Our observations yield $\delta\nu=58.58\pm0.02$\,GHz, clearly consistent with the former, and supported by the photometric redshift.

With the precise redshift and the observed $^{12}$CO line luminosities in
hand, we can estimate the molecular gas content of \smghy.
The observed global $^{12}$CO(5--4) to $^{12}$CO(3--2) line ratio (0.67$\pm$0.09) implies that
the CO emission is sub-thermally excited, at least for the CO(5--4)
line. The average line ratio reported by Bothwell et al.\ (2013) for a sub-sample of their 30 SMGs is r$_{53}$=0.62. This line ratio is similar to that observed for
SMM\,J16359+6612 (Weiss et al.\ 2005) and we use a similar scaling to estimate a $^{12}$CO(1--0) line luminosity of  $L'\approx1.86\cdot10^{11}$\,K\,\kms\,pc$^2$.  This translates into a
molecular gas mass of $M_{\hh}\approx2.1\cdot10^{11}\,\msol$ using a
standard ULIRG conversion factor of 1.0\,\msol\,(K\,\kms\,pc$^2$)$^{-1}$ found for typical $z\sim2.5$ SMGs in Bothwell et al.\ (2013), and including a 15\% contribution from Helium.
  (Downes et al.\ 1998 adopted 0.8\,\msol\,(K\,\kms\,pc$^2$)$^{-1}$).

\subsection{SED modelling}

The SED of \smghy, with the Swinbank et al.~(2014) composite SMG template overlaid is shown in Fig.~4. Significantly hotter or cooler SEDs than this $\sim35K$ template are reasonably rejected by the photometry. The template is split at a rest frame 10$\mu$m in order to fit the optical/near-IR faint component as decoupled from the luminous far-IR to submm dust component.  
The far-IR luminosity of this model is $3.8\times10^{13}\,\lsol$, translating to a SFR of 2500 M$_\odot$ yr$^{-1}$ (Kennicutt 1983).
\smghy\ is not unusually faint in optical/IR magnitudes for SMGs with S$_{\rm 850\mu m}$$\sim$3-5\,mJy, but its far-IR output is significantly higher, and it is thus proportionally far more obscured than ALESS SMGs (see Swinbank et al.\ 2014).

A hyper-$z$ fit to the IRAC and near-IR photometry yields a stellar mass of 3$\times10^{11}$ M$_\odot$, assuming a Chabrier IMF.  Together with the above gas mass estimate,  the gas fraction is $\sim$40\%, and precludes using substantially higher conversion factors, $\alpha$(CO)$\gg1$, which would drive the gas fraction up towards 100\%.

\subsection{SMA and PdBI continuum maps and source multiplicity} % study of flux and source size}

The PdBI  observations reveal only a single significant detection in both continuum and line emission, showing that there is no obvious multiplicity at 2mm/3mm or in CO lines.  As such there are no nearby ($<25''$ primary beam) $^{12}$CO-luminous companions to \smghy.
Similarly the SMA image does not detect any additional significant sources.
The combined 3mm continuum map reaches 0.35\,mJy rms. At this depth we are sensitive to SMGs with 
SFRs$\sim$1000\,M$_\odot$ yr$^{-1}$. The 2mm continuum map reaches  1.2\,mJy rms and should detect similar SFRs for a typical $\beta\sim1.5$ dust SED. The SMA 870$\mu$m map rms of 1.1\,mJy is much more sensitive to dusty sources, and should detect galaxies with SFRs$\sim$300\,M$_\odot$ yr$^{-1}$.
We note that a CO-luminous DRG in this field with a SFRs$\sim$200\,M$_\odot$ yr$^{-1}$ from Chapman et al.\ (2015) is not detected in any of these continuum maps, nor are any of the other known  galaxies lying within the primary beam.

By considering only the SMA compact configuration data (2$''$ beam size), we measure a larger 
flux density of 19.3$\pm$2.8\,mJy, close to that measured by SCUBA-2 in the 14.5$''$ beam, although the large measurement error agrees with the extended configuration flux measurement within $\sim$2$\sigma$. 
Nonetheless, this may suggest
 there is either a modest conflict in the flux scale between compact and extended configurations, or that the source may be marginally 
 resolved on scales greater than the average beam size achieved in our extended configuration data, $>0.85''$. However,  this is unusual for bright SMGs (Simpson et al.\ 2015a) which typically are resolved with  $\sim0.3-0.4$ source sizes.
 Ubiquitous SMG multiplicity results from surveys (e.g., Wang et al.\ 2010, Hodge et al.\ 2013, Simpson et al.\ 2015b) and theoretical arguments (e.g.\ Hayward et al.\ 2012, 2013a, 2013b) would argue that there may well be faint companions within the SCUBA-2 beam, and the similarity of the compact SMA data flux may just be a lucky coincidence.

\section{Discussion}

\subsection{Gas dynamics}

At first glance the double %horn  
peaked line profile 
suggests a circumnuclear molecular toroid. However, if such a stable molecular gas distribution would supply
the large star formation rate ($\approx2500$\,\msol\,yr$^{-1}$),
we would expect a high gas excitation. The observed line ratio  r$_{53}$=0.67$\pm$0.09 is low compared to well-studied nearby starburst galaxies NGC\,253 and M\,82, although it is  typical of that measured by Bothwell et al.\ (2013), a brightness temperature ratio r$_{53}$=0.62,  for SMGs with  characteristic L$_{850\mu m}$ lower than \smghy\ by a factors of 3$\times$ on average.
However, the two line peaks could also be two disk components merging,  individually unresolved. %More telling is t
The line integrated gas properties between the spectral components are   similar in the red and blue component line ratios (r$_{\rm 53, red}$=0.6$\pm$0.1, r$_{\rm 53, blue}$=0.7$\pm$0.2), contrasting the component line widths and central velocities which do differ between the different excitation lines. This argues that the (5--4) line may be coming from a different volume of gas than the (3--2) line, which could be true in either a disk-like or merger configuration. If the SMA dust continuum size is the same as the CO gas ($<0.85''$ or $<6.7$ kpc), then either the disk or the merger is in a very compact state. %These points also argue for two distinct merging galactic components. 
We note that a ring-like rotating disk
distribution of gas was strongly endorsed as an explanation for another similar double-peaked SMG, SMM\,J02399-0136   (Genzel et al.\ 2003), 
however from multi-frequency, high-resolution line mapping observations, Ivison et al.\ (2010) showed that the system comprises a multiple merger. %cannot explain the differing excitation conditions in the two components.

\smghy\ is clearly a very massive system, and is in particular a very massive 
{\it baryonic} system.  Observational studies of local ellipticals (Keeton 
2001; Boriello, Salucci, \& Danese 2003; cf. Loewenstein \& Mushotzky 2003) 
suggest that typically $\sim 30\%$ of the total mass within the optical 
effective radius will be dark; the interior regions of massive local disk 
galaxies have similarly low dark matter fractions (Combes 2002).  By analogy 
with these local systems, \smghy\ should have a baryonic mass of $\geq 4 \times 
10^{11}\,{\rm sin}^{-2}\,i^{-1}$\,M$_\odot$ within $7\,
{\rm kpc}$; M$_{\rm gas} \sim 2\times10^{11}$\,M$_\odot$ 
then implies that a  
 limit to the stellar mass within the same radius is $\sim2\times10^{11}$\,M$_\odot$.  
The  Kennicutt (1983) %\citet{kenn83} 
stellar initial 
mass function (IMF) models of Cole et al.\ (2001) % \citet{cole01}
%\footnote{The \citet{kenn83} IMF 
%is comparable in this context to a \citet{salp55} IMF from $m_{\rm lower} = 
%1\,M_\odot$ to $m_{\rm upper} = 100\,M_\odot$.  For a $0.1-100\,M_\odot$ 
%Salpeter IMF, \citet{cole01} find $m^*(z = 0) = 1.4 \times 
%10^{11}\,M_\odot$.} 
imply that M$^* \sim 7 \times 
10^{10}$\,M$_\odot$ for the galaxy stellar mass function in the 
local Universe.  Based on the stellar content of its inner 
$7\,{\rm kpc}$, then, 
\smghy\ is already at least a $3$M$^*$ system, on its way to becoming a $\geq 
4$M$^*$ system as it turns the remainder of its molecular gas into stars.

\subsection{An Eddington-limited starburst?}

%dynamical mass to estimate the sigma value, Mdyn \propto sigma^2
 % Mdyn [M?] = 1.56 × 106 ?2 R,
%but you can also use a disk estimator for Mdyn which is 
%Mdynsin2i[M?]=4×104 V2 R, 

If we assume that \smghy\ is starburst dominated and make assumptions about the morphology and kinematics, we find it may be radiating close to or at the Eddington limit of its starburst.
Elmegreen et al.\ (1999) demonstrated  that feedback from ongoing star formation sets a physical limit on the minimum size of a star forming region. Radiation pressure from high luminosity star formation regions should produce strong momentum--driven winds due to high dust opacity seen by the ultraviolet light produced by young stars
(e.g., Netzer \& Elitzur\ 1993). These winds are confined by the gravitational potential, which scales as $\Phi \sim f_g \sigma^2 \log{D}$  for an isothermal sphere  (with the gas fraction $f_g$,   stellar velocity dispersion $\sigma$, diameter of the starburst region $D$).  In the optically thin limit (appropriate for optically thick clouds with a small volume filling factor embedded in the diffuse interstellar medium) and assuming a Chabrier et al.\ (2003) 
initial mass function (IMF)
this leads to a maximum star formation rate
\begin{equation}
{\rm SFR_{max}} = 360\, \sigma_{400}^2 D_{\rm kpc} \kappa_{100}^{-1}\,\,\, {\rm M}_{\odot}\,\,{\rm yr^{-1}}
\end{equation}
We measure the scale size of the starburst $D_{\rm kpc}$ as the Gaussian FWHM of the SMA detected source,
and $\sigma_{400}$ as the line--of--sight $^{12}$CO gas velocity dispersion in units of 400 km s$^{-1}$ which is the best fit single Gaussian  to the entire CO(3--2) line ;
$\kappa_{100}$ is the dust opacity in units of 100 cm$^{2}$ g$^{-1}$, which we set to 1,  but note that many dust models allow for significantly  higher opacity, and in particular for the ultraviolet radiation produced by young massive stars during a starburst (e.g., Li et al.\ 2001).    
\smghy\ has a very high far-IR luminosity, with L$_{\rm FIR} \approx 4\times 10^{13}$\, L$_\odot$ and ${\rm SFR} \approx 2500$ M$_\odot$ yr$^{-1}$ on  a physical scale of $<7$ kpc.  
This very luminous SMG is therefore near the Eddington limit $\leq2000$ M$_\odot$ yr$^{-1}$ for a starburst on those scales.  
The size/SFR ratio of \smghy\ is   typical of the more luminous SMG components identified with ALMA observations by Simpson et al.\ (2015a), suggesting near Eddington-limited starbursts are not unusual in the SMG population. This motivates obtaining additional higher resolution 870$\mu$m and $^{12}$CO observations of 
\smghy\ to truly resolve the emission of this bright galaxy.

\begin{figure}
\centering
\includegraphics[width=8.5cm,angle=0]{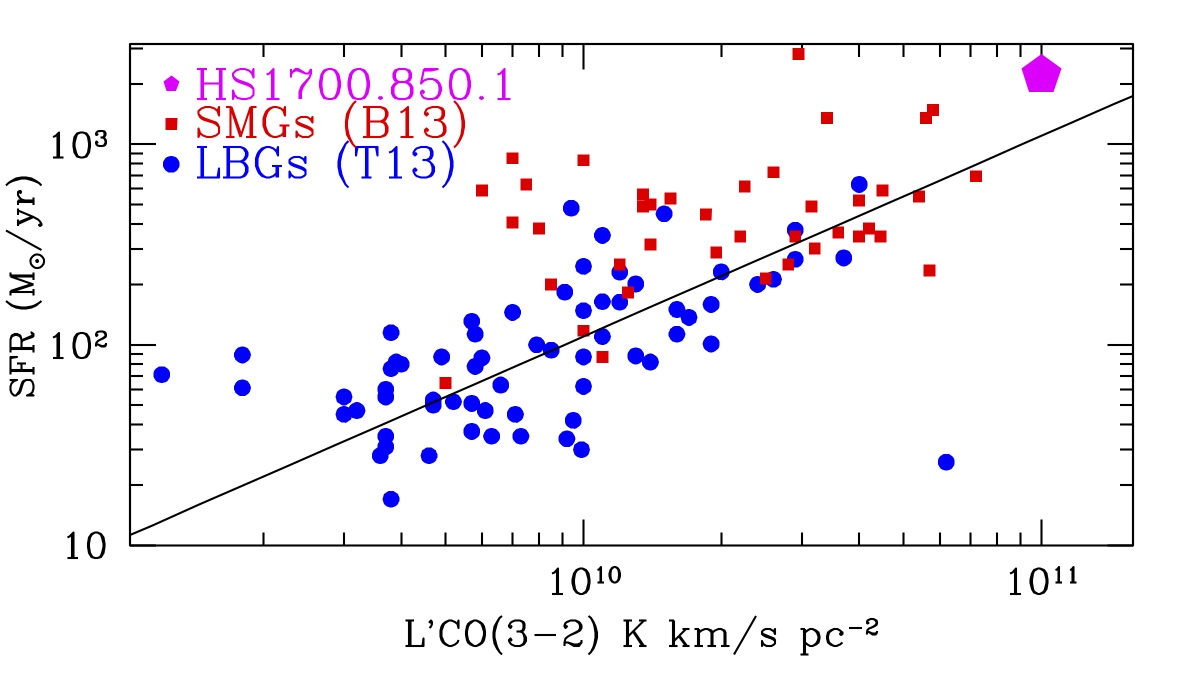} 
\caption{
SFR versus L$'$CO(3-2)   for various literature galaxies compared with \smghy\ (normal star forming galaxies -- Tacconi et al.\ 2013;  SMGs -- Bothwell et al.\ 2013). CO(3-2) is adopted for maintaining all displayed galaxies as close to observed transitions as possible. We  show L$'$(CO) for \smghy, revealing it is nominally under-luminous in $^{12}$CO compared to other ultra-luminous {\it normal} star forming galaxies, but not unusual compared to many SMGs in Bothwell et al.\ (2013) in lying a factor 2-3$\times$ above the Tacconi et al.\ (2013) best fit relation. 
}
\label{dustsed} 
\end{figure}

\subsection{L$'_{\rm CO}$ to L$_{\rm FIR}$ comparison with literature}

We next turn to a discussion of \smghy\ versus other %proto-cluster galaxies and versus other 
$^{12}$CO-detected sources, 
%{\bf LCO vs LFIR plot}
plotting L$'_{\rm CO}$-SFR for the known and published high-$z$ sources (Fig.~5), and translating L$_{\rm FIR}$ to SFR in the SMG populations using standard conversions (Kennicutt 1983). 
Of note is that \smghy\ has a significantly higher L$'_{\rm CO}$ than any of the 35 unlensed SMGs in the Bothwell et al.\ (2013) survey. 
Comparably $^{12}$CO-luminous SMGs were found in a few cases -- Ivison et al.\ (2013) identified a binary HyLIRG merger from a $\sim$200 deg$^2$ {\it Herschel} survey, while one lensed SMG from the $\sim$1300 deg$^2$ South Pole Telescope survey appears to be this luminous in $^{12}$CO after  a demagnification factor of $\sim5\times$ (Hezaveh et al.\ 2013).  Clearly finding SMGs with such high L$'_{\rm CO}$  has not been an easy task,  even probing the widest field mm-wave surveys.

 \smghy\ is somewhat under-luminous in $^{12}$CO for its SFR (or a high efficiency star former) compared to the average relation from Tacconi  et al.\ (2013) for {\it normal} star forming galaxies (shown in Fig.~5). However, compared to SMGs from Bothwell et al.\ (2013), \smghy\ is not unusual in lying 2-3$\times$ above the relation, an observation that supports application of the ULIRG conversion rate from L$'_{\rm CO}$ to total molecular gas mass.

\subsection{The environment of HS1700.850.1}

Given the extreme nature of \smghy, we  explore whether the surroundings of this SMG are unusually clustered.
The environment of HS1700.850.1 is traced in great detail by  the LBG survey in the field, presented in Fig.\ 6 as a
histogram binned to 500 km s$^{-1}$  to highlight any possible neighbouring galaxies. HS1700.850.1 appears to reside in a sparse region, with the closest (and only) neighbouring galaxy within $\pm$1500 km s$^{-1}$ at $z$=2.807 and lying  5$'$ (2.5\,Mpc) away. The LBG  selection function appropriate to this field (see Steidel et al.\ 2014) suggests that only two  LBGs are expected in a 1500  km s$^{-1}$ region from the 243 spectroscopic LBGs with measured redshifts  in  this field, thus finding only one about \smghy\ is not statistically significant.  There is one modest over density of six sources   at $z=2.751$, which includes the QSO HS1700+64 itself, contrasting with the sparse \smghy\ environment nearby.
Off the figure, at $z=2.31$, there is a large over-density consistent with a forming proto-cluster (Steidel et al.\ 2005).

Using the extensive spectroscopic database of the KBSS survey, Trainor \& Steidel  (2012) study similar extreme sources to \smghy\ at $z$=2.5-3.5 -- HyLIRG quasars (HLQSOs). They find 
the HLQSOs themselves are associated with a $\delta$$\sim$7 overdensity in redshift when considered on scales of $\sim$5 $h^{-1}$ Mpc,  corresponding to %. This overdensity has
a velocity scale of $\sigma_{\rm v,pec}$ = 200 km s$^{-1}$ %after subtracting the effect of redshift errors, 
and a projected scale of R$\sim$200 proper kpc. By studying the redshifts of galaxies around each HLQSO,
they find however that there are no overdense peaks of similar significance. 
Repeating this procedure  
on random galaxy redshifts shows that the HLQSOs are correlated with much more significant small-scale overdensities than the average galaxy.
They deduce that each HLQSO in their sample inhabits a DM halo with mass $\log$(M$_{\rm h}$/M$_\odot$) $>$ 12.1$\pm$0.5, or a median halo mass of $\log$(M$_{\rm h, med}$/M$_\odot$) = 12.3$\pm$0.5. 
Crucially, the number density of these halos exceeds the number density of HLQSOs by a factor $\sim10^6-10^7$, prompting the conclusion that the host halos of HLQSOs are not rare. 
Clearly \smghy\ follows this trend, inhabiting an environment that is indistinguishable from the average field galaxy density in Fig.~6.

\begin{figure}
\centering
\includegraphics[width=9.29cm,angle=0]{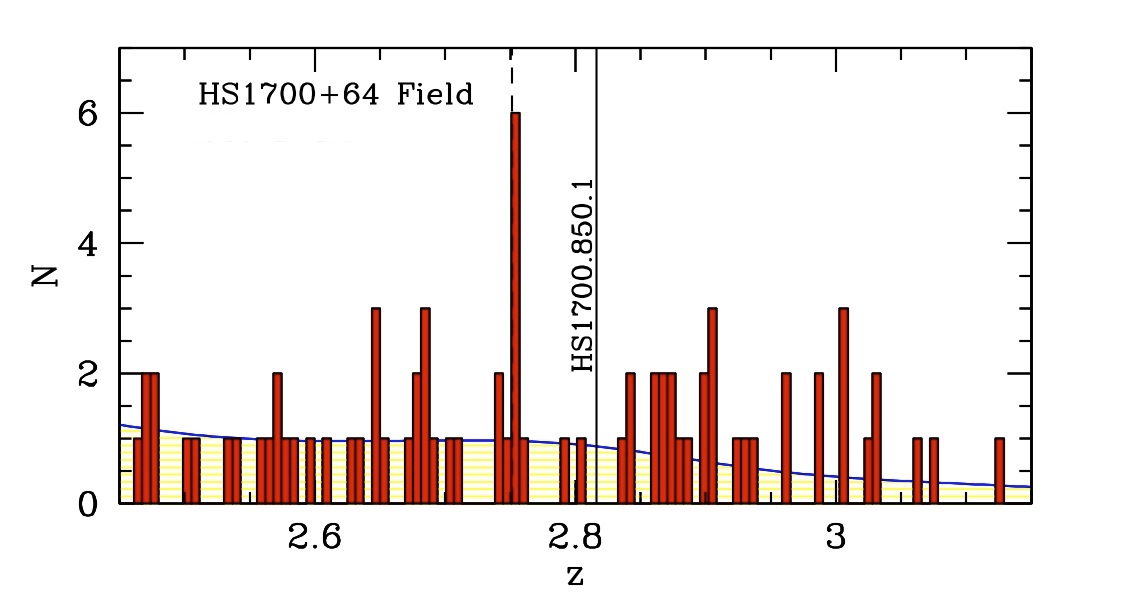}
\vskip-0.4cm
\caption{The environment of HS1700.850.1 derived from the LBG survey in the field, with 500 km s$^{-1}$ bins to highlight any possible neighbouring galaxies (the closest one at z=2.807 lies 5$'$ away). HS1700.850.1 appears to reside in a relative void, although the selection function suggests only two expected LBGs in a 1500  km s$^{-1}$ region for the 243 spectroscopic galaxies from $z=1.5$--3.5 measured in this field (the overall
redshift selection function for the continuum-selected redshift survey in
all 15 fields of KBSS is shown as a blue line -- see e.g.\ Steidel et al 2014).  There is a  6-object  galaxy spike at the redshift of the QSO HS1700+64 itself, contrasting the sparse environment of \smghy, while at $z=2.31$ (off the plot axes) there is in fact a very large galaxy over-density consistent with a forming proto-cluster (Steidel et al.\ 2005).
}
\label{spectra} 
\end{figure}

The scarcity of HLQSOs  is likely due to an extremely improbable small-scale phenomenon that produces HLQSOs, possibly related to an atypical galaxy interaction geometry or similar scenario, or else  extremely short durations of the events.
Analogously in \smghy\ and other extreme SMGs,
an unlikely merger trajectory perhaps with unusually gas-rich progenitors (e.g., Narayanan et al.\  2010) may explain rare bursts approaching S$_{\rm 850 \mu m}\sim$20 mJy and L$_{\rm IR}\sim2\times10^{13}$ L$_\odot$ (e.g., Ivison et al.\ 2013). This motivates obtaining more detailed followup of the resolved double-peaked $^{12}$CO emission in \smghy\ to better understand the circumstances generating such enormous star formation rates and illuminated dust masses.
We might also hypothesize that a very short duration starburst could explain the extreme luminosity of \smghy. 
For example, if we assumed that every S$_{\rm 850 \mu m}$=15-20\,mJy SMG was a brief episode in the more numerous 5\,mJy SMG population (1000$\times$ more numerous in the ALMA-based counts of Simpson et al.\ 2015b, but still very luminous starbursts in themselves), then the HyLIRG burst would last only 50\,kyr of a characteristic burst of 50\,Myr (e.g., Chapman et al.\ 2005, Tacconi et al.\ 2006). This may not be a well motivated scenario -- such a short  timescale   could likely only be driven by intense, hot AGN activity, yet there is no sign of an AGN in \smghy. Further, an AGN could not boost the 850$\mu$m emission while keeping the typical SMG appearance in Fig.~4.
Most likely the SMG HyLIRGs are driven by a combination of factors.

Another estimate on the halo mass can be obtained using dynamical mass estimates from the peculiar velocities and the projected scale of the galaxy overdensities. With this approach  Trainor \& Steidel (2012)  estimate  HLQSOs are on average associated with even larger group-sized environments with total mass $\log$(M$_{\rm grp}$/M$_\odot$)$\sim$13.  The SMG clustering analysis in Hickox et al.\ (2012) suggested that the  SMG population is hosted by haloes of comparable mass (see also Blain et al.\ 2004).
This overdense environment with small relative velocities would increase the probability of a rare event, but an unusual merger configuration is still likely required to generate such large  luminosities.
In \smghy, even a group environment of this mass is unlikely  based on the lack of explicit companions within 1500 km s$^{-1}$ and within $\sim$2.5 Mpc. However, a  $\sim10^{13}$ M$_\odot$ halo will typically only have a couple of bright LBG-class galaxies in it, and might be easy to miss in an incomplete spectroscopic survey.

%In spite of the extremely large black hole masses implied by the observed luminosities of HLQSOs [log(MBH/M$_\odot$) $\sim$ 9.7], 
The Trainor \& Steidel (2012) analysis has  clearly  demonstrated  that HyLIRG-class sources do not require environments very different from their much less luminous counterparts. Evidently, the exceedingly low space density of at least HLQSOs ($\sim$10$^{-9}$ Mpc$^{-3}$) results from a one-in-a-million event on scales $\ll$ 1 Mpc, and not from being hosted by rare dark matter halos.
\smghy\ appears to inhabit a similarly typical halo.
A moderate mass environment  was also noted for a highly significant association of 8 SMGs at $z$$\sim$1.99 by Chapman et al.\ (2009), whilst revealing that $z$=2--3.5 SMGs generally do not align well with LBG overdensities within the GOODS-N survey field. Similar anti-biasing of SMGs is noted in the ECDFS field (A.\ Danielson in preparation).  Notable exceptions exist at higher redshifts, including  a $z$=4.05 galaxy overdensity found through an association of bright SMGs (Daddi et al.\ 2009),  the COSMOS $z$=5.1 structure (Capak et al.\ 2011), and HDF850.1, which apparently lies in a $z$=5.3 galaxy overdensity (Walter et al.\ 2012).

%Miller et al.\ (2015)  used the Bolshoi simulation (e.g., Behroozi, Wechsler \& Conroy 2013) and various toy models for the SMG population 
Miller et al.\ (2015) used the Hayward et al.\ (2013a) model for the SMG population to estimate typical environments of the brightest SMGs and associations of multiple SMGs found in $\sim$500 Mpc$^3$ volumes. This model is based on mock galaxy catalogues derived from the Bolshoi simulation (Klypin et al.\ 2011; Behroozi et al.\ 2013) and insights obtained from performing radiative transfer on hydrodynamical simulations (Hayward et al.\ 2011; Hayward et al.\ 2013b).
Miller et al.\ (2015) find the SMGs inhabit a range of mass scales going down to the average cosmic density in the simulation, and rarely are found in the most massive regions at these $z=2$--3 epochs (although at significantly higher redshifts they are better tracers of the most massive regions). The surrounding LBG survey indeed suggests \smghy\ is statistically consistent with inhabiting the average cosmic density.

\section{Conclusions}

We  presented an IRAM-PdBI spectral survey of the HyLIRG \smghy, the brightest 850$\mu$m source found in the SCUBA-2 survey of the HS1700+64 field, and indeed all six SCUBA-2/KBSS  fields. Our primary goal was to identify its redshift, guided by photometric redshift estimates in the optical through far-IR, and study its properties and environment.
We conclude: 

$\bullet$ The redshift for \smghy\ is $z$=2.816, secured through the  $^{12}$CO(3--2) and (5--4) lines, validating the use of crude photometric redshifts to guide $^{12}$CO redshift searches.

$\bullet$ The clear doubled-peaked profile of \smghy\ in both CO transitions is likely a result of an end stage merger, within 6\,kpc, as constrained by the size of the 850$\mu$m emission. 

$\bullet$ \smghy\ exhibits typical L$'_{\rm CO}$ for its far-IR luminosity  or SFR compared to other unlensed SMGs, although its L$'_{\rm CO}$ is significantly larger than any SMGs identified in the Bothwell et al.\ (2013) survey.

$\bullet$ The environment of \smghy\ is indistinguishable from the average field galaxy density, and no nearby LBG companions are identified within 1500 km s$^{-1}$. The HyLIRG burst could result from a 10$^{-6}$ unlikely event on scales $\ll$ 1 Mpc, and not from being hosted by a rare dark matter halo, but could also be due to a very short $\sim50$\,kyr burst timescale.

\acknowledgments 

We thank an anonymous referee for constructive comments on the manuscript.
IRAM is supported by INSU/CNRS (France), MPG (Germany) and IGN (Spain). 
The Submillimeter Array is a joint project between the Smithsonian Astrophysical Observatory and the Academia Sinica Institute of Astronomy and Astrophysics and is funded by the Smithsonian Institution and the Academia Sinica.
We thank summer students K.\ Lacaille and R.\ Perry for their work on the KBSS-SCUBA2 survey project.
SCC acknowledges NSERC and CFI for support.
FB acknowledges support through the Collaborative Research Centre 956, sub-project A1, funded by the Deutsche Forschungsgemeinschaft (DFG).
IRS acknowledges support from STFC, a Leverhulme Fellowship, the ERC Advanced Investigator programme DUSTYGAL 321334 and a Royal Society/Wolfson Merit Award. 
JEG acknowledges support from the Royal Society.

%%%%%%%%%%%%%%%%%%%%%%%%%%%
% BIBLIOGRAPHY
%%%%%%%%%%%%%%%%%%%%%%%%%%%

\end{document}